\documentclass[12pt]{article}
\usepackage{latexsym,amsfonts}

\def\ith{i^{\underline{{\rm th}}}}

\def\bear{\begin{array}}
\def\bun{{\cal V}}
\def\eear{\end{array}}
\def\Proj{{\rm Proj}}

\def\tl{\tilde\lambda}

\def\gso{{\mathfrak{so}}}

\def\beq{\begin{equation}}
\def\eeq{\end{equation}}

\newcommand{\nnn}[1]{(\ref{#1})}

\newcommand{\tfrac}[2]{{\textstyle{\frac{#1}{#2}}}}

\def\eye{\sqrt{-1}}

\let\a=\alpha
\let\b=\beta

\let\d=\delta
\let\D=\Delta
\let\e=\varepsilon
\let\f=\varphi

\let\i=\iota

\let\l=\lambda
\let\L=\Lambda
\let\m=\mu

\let\N=\nabla
\let\nd=\nabla

\let\r=\rho

\let\w=\Upsilon
\let\W=\Omega

\newcommand{\V}{{\mbox{\sf P}}}
\newcommand{\J}{{\mbox{\sf J}}}

\def\cA{{\cal A}}
\def\cB{{\cal B}}

\def\cE{{\cal E}}
\def\ce{\cE}
\def\co{{\cal O}}

\def\cW{{\cal W}}

\newtheorem{theorem}{Theorem}

\newtheorem{corollary}[theorem]{Corollary}

\newcommand{\nn}[1]{(\ref{#1})}

{\hfill $\Box$ \end{trivlist}}

\begin{document}

\title{A conformally invariant differential operator on Weyl tensor densities}
\author{Thomas Branson and A. Rod Gover}
\date{}

\maketitle

\begin{abstract}We derive a tensorial 
formula for a fourth-order conformally
invariant differential operator on conformal
4-manifolds.  This operator is applied to algebraic Weyl tensor
densities of a certain conformal weight, and takes its values in
algebraic Weyl tensor densities of another weight.  For oriented
manifolds, this operator reverses duality: For example in the
Riemannian case, it takes self-dual to anti-self-dual tensors and vice
versa.  We also examine the place that this operator occupies in known
results on the classification of conformally invariant operators, and
we examine some related operators.\end{abstract}

\subsection{Introduction}\label{intro}

Recent work on anomalies in conformal field theory \cite{des}
has revealed a potentially important role
for a certain conformally
invariant linear differential operator $ D$ 
in dimension 4.  This operator
has order 4, and acts on tensor-densities of the symmetry and trace type 
of the Weyl conformal curvature tensor.
The output of this operator is a tensor-density of a different
conformal weight, but also
of the symmetry and trace type of the Weyl tensor.
Under this operator, self-dual and anti-self-dual
Weyl tensor densities are interchanged, in a way reminiscent of the
chirality switch effected by the Dirac operator, and the duality
switch effected by the 
the middle-form-density 
operator $\d d-d\d+$(Ricci correction) of \cite{tbcpde}. 

The existence of this operator is probably first due to Eastwood and
Rice \cite{ER}. Their work constructed a very large class of invariant
differential operators on conformal 4-manifolds, and in the process,
pioneered an approach now known as the {\em curved translation
principle}.  This technique has since been developed significantly,
and for conformal manifolds of any dimension $n\geq 3$, many
differential operator existence questions can be settled by consulting
\cite{ES}.  
(For a recent complete treatment of large classes of 
invariant operators in the setting of general
parabolic geometries,
see \cite{CSSann}.)
However, even given the existence of a particular
operator, producing a useful and explicit formula is sometimes a
non-trivial matter. In \cite{goox,Gothesis} formulas for the operator
$D$, as well as many of the related operators discussed below, are
obtained by a rather different construction which uses ideas from a
twistor theory.  In fact, there are universal formulas which yield $D$
and many of its relatives; a principle of this type is formulated in
Theorem \ref{lambdathm} 
below, which may be viewed as an elementary exposition of
a class of special cases of the general results of
\cite{goox,Gothesis}.  We discuss the universal formulas and general
results in Section \ref{standard} below.  
More recently, universal
formulas along these lines have been recovered in an even more general
setting in \cite{CSSstandard}, this time via a construction which
explicitly uses the normal Cartan connection associated to a 
parabolic
geometry.  

In Corollary \ref{weyltwo} below, we take the Weyl tensor density operator
that motivated the present work and make it even more readily usable,
by giving a formula for it in standard abstract index notation.
Essentially, this explicitly accomplishes the projections involved
in formulas like that of our Theorem \ref{lambdathm}.

In the construction leading up to Theorem \ref{lambdathm}, we show how
formulas for high-order invariant operators can be built 
using information about first-order invariant operators;
in this case the {\em Stein-Weiss operators} or {\em generalized
gradients} of \cite{sw,feg}.

The authors gratefully 
acknowledge support from US NSF grant INT-9724781.

\subsection{Preliminaries}\label{prelim}

We shall work for now in the setting of Riemannian conformal geometry.
Many of our ultimate conclusions about the existence of invariant
operators on tensor-densities and their abstract index formulas will,
however, be independent of the metric signature.  We shall take stock
of this in Section \ref{epilog} below. 

Natural irreducible tensor
bundles in oriented $n$-dimensional Riemannian
conformal geometry are labelled
by a dominant SO$(n)$-weight $\l$ and a conformal weight $w$; we
shall write such labels in the form $[w|\l]$.  
The parameter $w$
is a real number, the {\em density weight}, and $\l$ is an
$\ell:=[n/2]$-tuple of integers satisfying the {\em dominance} condition
\beq\label{dom}
\bear{ll}
\l_1\ge\l_2\ge\cdots\ge|\l_\ell| & (n\ {\rm even}), \\
\l_1\ge\l_2\ge\cdots\ge\l_\ell\ge 0\qquad & (n\ {\rm odd}).
\eear
\eeq
Another important label is the {\em rho-shift} of $[w|\l]$:
\beq\label{round}
\bigl[\bigl[\tilde w|\tl\bigr]\bigr]=
\bigl[\bigl[w+\tfrac{n}2\big|\tl+\r_{\gso(n)}\bigr]\bigr],
\eeq
where 
$$
\r_{\gso(n)}=\left(\tfrac{n-2}2,\tfrac{n-4}2,\ldots,\tfrac{n-2\ell}2\right).
$$
We use the extra set of brackets in \nnn{round} advisedly,
as a reminder of whether we have or have not rho-shifted.
The string to the right of the bar in a rho-shifted label is
{\em strictly dominant}, that is the $\ge$ signs in \nnn{dom} are replaced
by $>$ signs.

Let $\bun[w|\l]$ or $\bun[[\tilde w|\tl]]$ denote the bundle with the
given label.  Then, for example, the conformal Laplacian (Yamabe operator)
$L=-\nd^a\nd_a+(n-2)R/(4(n-1))$ carries 
$\bun[\tfrac{2-n}2|0,\ldots,0]$ to $\bun[\tfrac{-2-n}2|0,\ldots,0]$
in a conformally invariant way: changing the metric $g$ to 
$\hat g=\W^2g$, where $\W$ is a positive smooth function, has no
effect on the operator.  If we force the operator to act between
bundles of the ``wrong'' density weights, 
we get an operator which is {\em conformally covariant} instead of
{\em invariant}.  For example, if we view the Yamabe operator 
as carrying
$\bun[0|0,\ldots 0]$ to $\bun[0|0,\ldots 0]$,
then replacement of 
$g$ by $\hat g$
gives an operator 
\beq\label{yamcov}
\hat Lf=\W^{-(n+2)/2}L(\W^{(n-2)/2}f)
\eeq
on
smooth functions $f$.
The concept of conformally covariance (as opposed
to invariance)
is useful, for example, when one wishes to 
have a spectrum.

If a metric is specified, i.e.\ if we are in the setting of
Riemannian geometry, irreducible tensor bundles are parameterized
simply by the $\l$ above.
We shall denote by $\bun(\l)$ the bundle with
the given (non-rho-shifted) label.

There is a chance of having a conformally invariant operator
$\bun[[w|\l]]\to\bun[[w'|\l']]$ only if the length $\ell+1$ strings
$(w,\l)$ and $(w',\l')$ are related by
$$
\bear{ll}
\mbox{a permutation and an even number of sign changes},\qquad 
& n\ {\rm even}, \\
\mbox{a permutation and any number of sign changes}, & n\ {\rm odd}.
\eear
$$
That is, the rho-shifted weights $[[w|\l]]$ and $[[w'|\l]]$ must be
similar under the {\em affine Weyl group}.
(Dually this is equivalent to the corresponding generalized Verma
modules having the same 
{\em central character} for the enveloping
algebra of ${\mathfrak{so}}(n+2,{\mathbb{C}})$.
See \cite{ES} for further details on this and related points here.)
Even on round $S^n$, this is a necessary condition 
for a nontrivial differential operator 
invariant under the group of conformal diffeomorphisms.
An additional necessary condition is that the pair
$[[w|\l]]$ and $[[w'|\l']]$
have one of the correct relative placements in the 
{\em Bernstein-Gelfand-Gelfand} diagram made from the affine
Weyl orbit of $[[w|\l]]$.

Of the differential operators on round $S^n$ that are invariant 
under the conformal group, all are 
known to have invariant
generalizations to arbitrarily curved manifolds, except the
{\em longest arrows} in even dimensions $n\ge 4$ -- 
operators carrying $[[u|\m]]\to[[-u|\bar{\m}]]$, 
where $u>\m_1$ and $\bar{\m}=(\m_1,\ldots,\m_{\ell-1},-\m_\ell)$.
(See \cite{ES}.) 
These generalizations need not be unique, but they are differential
operator invariants of conformal structure which evaluate to 
the given (unique up to a constant factor) operator on round $S^n$.
An example of an even-dimensional longest arrow that 
{\em does} generalize is the
{\em Paneitz operator} $\bun[[2|1,0]]\to\bun
[[-2|1,0]]$ in dimension 4 (see \cite{pan}, \cite{riegert}, \cite{es}); 
or more generally
\cite{gjms},
the GJMS operator $P_n:[[\ell|\rho_{\gso(n)}]]\to[[-\ell|\rho_{\gso(n)}]]$
in even dimension $n$.
An example of one which does not generalize \cite{crgnonex} is the operator
with principal part $\D^3$ on scalar densities in $S^4$; here the labels
are $[[3|1,0]]\to[[-3|1,0]]$.

\subsection{A class of fourth-order conformally invariant operators}
\label{stdfourth}

If $\l$ is an $\ell$-tuple, let $\l_i$ be its $\ith$ entry.  
(Recall that $\ell$ is the integer $[n/2]$.) 
Let $e_i$ be the $\ell$-tuple with 1 in the $\ith$ slot and 0 elsewhere.
If $\tl\pm 2e_i$ are strictly dominant SO$(n)$-weights, then so
are $\tl\pm e_i$ and $\tl$.
Suppose we try to approximate the 
conformally invariant operator which carries
\beq\label{theop}
\bun[[-(\tl_i-2)|\tl+2e_i]]\to 
\bun[[-(\tl_i+2)|\tl-2e_i]] 
\eeq
by composing
operators
\beq\label{constituents}
\bear{l}\bun[[-(\tl_i-2)|\tl+2e_i]]\to\bun[[-(\tl_i-1)|\tl+e_i]]\to
\bun[[-\tl_i|\tl]] \\
{}\qquad\to\bun[[-(\tl_i+1)|\tl-e_i]]
\to\bun[[-(\tl_i+2)|\tl-2e_i]].
\eear
\eeq
This is the unique path composing four first-order Riemannian invariant
differential operators. 
However, it is not a composition of 
conformally invariant operators, since it is never
the case that all five bundles involved 
are in the same affine Weyl orbit.
However, by \cite{feg}
there are conformally invariant operators
\beq\label{fourops}
\begin{array}{rl}
D_1:\bun[[-(\tl_i+1)\,|\,\tl+2e_i]]&\to\bun[[-(\tl_i+2)\,|\,\tl+e_i]], \\ 
D_2:\bun[[-\tl_i\,|\,\tl+e_i]]&\to\bun[[-(\tl_i+1)\,|\,\tl]], \\ 
D_3:\bun[[-(\tl_i-1)\,|\,\tl]]&\to\bun[[-\tl_i\,|\,\tl-e_i]], \\ 
D_4:\bun[[-(\tl_i-2)\,|\,\tl-e_i]]&\to\bun[[-(\tl_i-1)\,|\,\tl-2e_i]].
\end{array}
\eeq
(See below for tensorial realizations of these operators in the 
case of the Weyl tensor density problem.)  
In fact, these are the Stein-Weiss {\em gradients}, or compressions
of the covariant derivative.  For example, 
$$
D_1={\rm Proj}_{\tl+e_i}\nd|_{\tl+2e_i}\,.
$$

An invariant operator
$D:\bun[[w|\tl]]\to\bun[[w'|\tl']]$, when realized as an operator
$\bun[[a|\tl]]\to\bun[[b|\tl']]$, has conformal variation
$$
\frac{d}{d\e}\Bigg|_{\e=0}D_{e^{2\e\w}g}=:D'(\w)=(b-w'-a+w)\w D-(a-w)[D,\w].
$$
For example, recall \nnn{yamcov} above.
The $\w$ in the operator commutator $[D,\w]$
is an abbreviation for the multiplication operator $\f\mapsto\w\f$.
 
Thus the conformal variation of the composition of the operators in
\nnn{fourops} is
\beq\label{firstvar}
\bear{l}
(D_4D_3D_2D_1)'(\w)=3D_4D_3D_2[D_1,\w]+D_4D_3[D_2,\w]D_1 \\
{}\qquad-D_4[D_3,\w]D_2D_1-3[D_4,\w]D_3D_2D_1.
\eear
\eeq
One can get a differential operator of homogeneity 4 
(i.e.\ one which is scaled by $\a^{-4}$ when the metric is scaled
by a constant $\a^2$)
from $\bun(\l+2e_i)$ to 
$\bun(\l-2e_i)$ 
because the cotangent bundle is SO$(n)$-isomorphic to $\bun(e_1)$, and
$(\otimes^4\bun(e_1))
\otimes\bun(\l+2e_i)$
has a copy of $\bun(\l-2e_i)$ in its SO$(n)$ decomposition.  In
fact, there is just a single copy, and it lives in the subbundle
$\cE_{(abcd)_0}\otimes\bun(\l+2e_i)$, where $\cE_{(a_1\cdots a_p)_0}$ 
is the trace-free
symmetric part of the fourth tensor power of the cotangent bundle.
(To see how this fits into our general notation for tensors and 
tensor densities,
see the beginning of Section \ref{tens} below.)

Note that $\cE_{(a_1\cdots a_p)_0}\cong_{{\rm SO}(n)}
\bun(pe_1)$, and we need to
drop four units in one of the entries to get from $\l+2e_i$ to
$\l-2e_i$.  Thus summands of $\otimes^4\bun(e_1)$ which are 
isomorphic to, for
example, $\bun(3e_1+e_2)$, cannot contribute.  Let us say two indexed
expressions $A$ and $B$ are {\em equivalent}, and write $A\sim B$, if
they have the same trace-free symmetric part in their free indices.
For example, $\nd_a\nd_b\nd_c\sim\nd_b\nd_a\nd_c$ and
$g_{ab}\nd_c\nd_d\sim 0$.  In particular, if we have a 4-index
expression $A$ which gives a differential operator from
$\bun(\l+2e_i)$ to $\bun(\l-2e_i)$ via
$\cA=\Proj_{\bun(\l-2e_i)}A|_{\bun(\l+2e_i)}$, 
then $A$ may be
replaced by any equivalent expression without affecting the value of
$\cA$.  
Applying this to the problem at hand, we get from \nnn{firstvar}
that 
$$
\bear{rl}
(D_4D_3D_2D_1)_{abcd}&\sim\nd_a\nd_b\nd_c\nd_d\,, \\
(D_4D_3D_2D_1)'(\w)_{abcd}&\sim 3\nd_a\nd_b\nd_c\w_d+\nd_a\nd_b\w_c\nd_d
-\nd_a\w_b\nd_c\nd_d \\
&\qquad\qquad-3\w_a\nd_b\nd_c\nd_d \\
&\sim 10\w_{ab}\nd_c\nd_d+10\w_{abc}\nd_d+3\w_{abcd}.
\eear
$$
Here and below, we abbreviate $\nd_b\nd_a\w$ as $\w_{ab}$, and similarly
for other strings of derivatives of $\w$.
This already tells us that the composition $D_4D_3D_2D_1$ is invariant
under the conformal transformation group of $S^n$, 
since the infinitesimal conformal factors of that group
(the homogeneous coordinate functions) have vanishing 
trace-free 
symmetrized
covariant derivatives of order 2 and higher.

In dealing with conformal variation, 
it is often convenient to decompose the 
Riemann curvature tensor into the Weyl tensor $C^a{}_{bcd}$
and a trace renormalization $\V_{ab}$ of the Ricci tensor:
$$
R_{abcd}=C_{abcd}+2g_{c[a}\V_{b]d}+2g_{d[b}\V_{a]c}\,.
$$
Part of the convenience of $\V_{ab}$ derives from its conformal
variational formula
$(\V_{ab})'(\w)=-\w_{ab}\,$.  Together with 
the above, this suggests trying to correct by
adding (inside the compression 
$\Proj_{\bun(\l-2e_i)}\cdot|_{\bun(\l+2e_i)}$)
\beq\label{ten}
10\V_{ab}\nd_c\nd_d+10(\nd_a\V_{bc})\nd_d+3(\nd_a\nd_b\V_{cd}).
\eeq
To compute the conformal variation of this, first note that
$$
\bear{rl}
(\nd_a\V_{bc})'(\w)&\sim-\w_{abc}-4\w_a\V_{bc}, \\
(\nd_a\nd_b\V_{cd})'(\w)&\sim-\w_{abcd}-4\w_{ab}\V_{cd}-10\w_a\nd_b\V_{cd}\,.
\eear
$$
In addition, the $\nd\nd$ and $\nd$ on the right of compositions in
\nnn{ten} may be replaced by (respectively) $D_2D_1$ and $D_1\,$.
The new expressions are not the same, but they are equivalent.
The upshot is that
the conformal variation of 
$$
\nd_a\nd_b\nd_c\nd_d+
10\V_{ab}\nd_c\nd_d+10(\nd_a\V_{bc})\nd_d+3(\nd_a\nd_b\V_{cd})
$$
is equivalent to $18\w_{ab}\V_{cd}$.  But this is equivalent to the
conformal variation of $-9\V_{ab}\V_{cd}\,$.  

If we wish to move in the other direction, from $\bun(\l-2e_i)$ to
$\bun(\l+2e_i)$, the calculation is similar; we just have to change
a few signs.

\begin{theorem}\label{lambdathm} Suppose $\l+2e_i$ and $\l-2e_i$
are dominant SO$(n)$-weights.  Then
the differential operators
\beq\label{projform}
\bear{l}
\Proj_{\bun(\l\mp 2e_i)}
(\nd_a\nd_b\nd_c\nd_d+
10\V_{ab}\nd_c\nd_d+10(\nd_a\V_{bc})\nd_d \\
{}\qquad+3(\nd_a\nd_b\V_{cd})
+9\V_{ab}\V_{cd})
|_{\bun(\l\pm 2e_i)}
\eear
\eeq
are conformally invariant $\bun[\mp\tl_i+2-\frac{n}2|\l\pm 2e_i]\to 
\bun[\mp\tl_i-2-\frac{n}2|\l\mp 2e_i]$. 
In particular, in dimension 4, if $\l_1\ge|\l_2|+2$, then 
there are invariant operators 
$\bun[\mp(\l_1+1)|\l_1\pm 2,\l_2]\to\bun[\mp(\l_1+1)-4|\l_1\mp 2,\l_2]$
and $\bun[\mp\l_2|\l_1,\l_2\pm 2]\to\bun[\mp\l_2-4|\l_1,\l_2\mp 2]$.
As a special case of this in dimension 4 {\rm(}with $\l=(2,0)${\rm)}, 
there are invariant operators
$\bun[0|2,\pm 2]\to\bun[-4|2,\mp 2]$.
\end{theorem}

\subsection{Tensorial realizations}\label{tens}
Let us now consider tensorial realizations.  
Let $\cE[w]$ be the
bundle of $w$-densities; this is a realization of $\bun[w|0,\ldots,0]$.
Tensor bundles will be denoted by adorning the symbol $\cE$ with
the index configuration of their sections; thus the tangent bundle
is $\cE^a$ and the cotangent bundle is $\cE_a$.  
Standard symmetry type notation will also be used, as for example
when we spoke of $\cE_{(abcd)_0}$ above, or as in the example of
the 
exterior 2-form bundle $\cE_{[ab]}\,$.  The 
tensor product with a density bundle will be abbreviated, for 
example, by $\cE_{[ab]}[w]:=\cE[w]\otimes\cE_{[ab]}\,$.
Because the conformal metric is an element of $\cE_{(ab)}[2]$,
the raising and lowering of indices has an effect on the weight.
For example, $\cE^a\cong_{{\rm CO}(n)}\cE_a[2]\cong_{{\rm CO}(n)}
\bun[1|1,0,\ldots,0]$,
and $\cE_a\cong_{{\rm CO}(n)}\bun[-1|1,0,\ldots,0]$, where
CO$(n)$ denotes the extension of the structure group SO$(n)$ by
pointwise scalings.
 
The Weyl tensor $C^a{}_{bcd}$ of the metric is conformally invariant,
and thus is a section of $\cW^a{}_{bcd}\,$, where we use $\cW$ to denote
curvature symmetries and the absence of traces.
By the above remarks on the tangent and cotangent bundles, 
$C^a{}_{bcd}$ also lives in a copy of
$$
\bun[1|1,0,\ldots,0]\otimes(\otimes^3\bun[-1|1,0,\ldots,0]).
$$
Thus the Weyl tensor is a section of 
a direct sum of irreducibles bundles having the form
$\bun[-2|\l]$.  In fact, it is a
section of $\bun[-2|2,2,0,\ldots,0]$ if $n\ge 5$, and to
$\bun[-2|2,2]\oplus\bun[-2|2,-2]$ if $n=4$.  
The two summands in dimension 4 correspond to the
two dualities, or eigenvalues of the Hodge $\star$ applied in the
$cd$ index pair; these will be denoted $\cW_\pm$.
Algebraic Weyl tensor-densities are obtained
by tensoring with density bundles.  Examples that
are relevant for what follows are
$$
\cW^a{}_b{}^c{}_d\cong_{{\rm CO}(n)}\cW^a{}_{bcd}[2]
\ \ {\rm and}\ \ 
\cW_{abcd}\cong_{{\rm CO}(n)}\cW^a{}_{bcd}[-2].
$$

The following is just the last conclusion of
Theorem \ref{lambdathm} stated in these terms:

\begin{corollary}\label{mainops} 
If $n=4$, formula 
\nnn{projform} for $\l=(2,0)$ gives conformally invariant operators
$D_\pm$ from 
$(\cW_\pm)^a{}_b{}^c{}_d$
to 
$(\cW_\mp)_{abcd}\,$.
\end{corollary} 

The following will make it clear that there
is a unified tensorial formula for $D_+$ and $D_-$, so that one need
not actually accomplish the decomposition
into self-dual and anti-self-dual parts in order to apply the formula for
the
operator.  That is, the tensorial formula we shall give is really one
for the operator which is, in block form,
\beq\label{block}
D=\left(\bear{cc}\ 0\ &\ D_-\ \\ \ D_+\ &\ 0\ \eear\right).
\eeq

To get our tensorial realization,
choose a metric $g$.
If $Q_{abcd}$ is a differentio-tensorial 
expression, for example $\N_a\N_b\N_c\N_d$ or $\V_{ab}\V_{cd}$, 
and $Y$ is an algebraic Weyl tensor,
we define
$$
\begin{array}{rl}
  (Q\bullet Y)_{abcd} &= Q_{(acef)_0}Y^e{}_b{}^f{}_d
                 -Q_{(adef)_0}Y^e{}_b{}^f{}_c \\
&\qquad                 +Q_{(bdef)_0}Y^e{}_a{}^f{}_c 
                 -Q_{(bcef)_0}Y^e{}_a{}^f{}_d.
\end{array}
$$
We claim that $Q\bullet$ is a nonzero 
SO$(n)$-equivariant action of $\cE_{(abcd)_0}$ on algebraic
Weyl tensors interchanging the self-dual and anti-self-dual summands.

First note that $Q\bullet$ propagates the curvature symmetries: if
$Y$ satisfies
$$
Y_{abcd}=Y_{cdab}=Y_{[ab]cd}=-Y_{acdb}-Y_{adbc}\,,
$$
then $Q\bullet Y$ behaves similarly.  
The statements on trace and duality follow from the fact that
\beq\label{tp}
\bear{l}
\bun(4,0)\otimes\bun(2,\pm 2)\cong_{SO(4)} \\
\bun(2,\mp 2)\oplus\bun(3,\mp 1)\oplus
\bun(4,0)\oplus\bun(5,\pm 1)\oplus\bun(6,\pm 2).
\eear
\eeq
In particular, the bundles of algebraic Weyl tensors on the right and
left sides have opposite duality.  Traces of $Q\bullet Y$ would need
to land in $\bun(2,0)\oplus\bun(1,1)\oplus\bun(1,-1)\oplus\bun(0,0)$,
none of whose summands occur on the right in \nnn{tp}.  And in fact,
it is easily computed that the $ac$-trace, and thus any trace, of
$(Q\bullet Y)_{abcd}$ vanishes.

To show that $Q\bullet$ is nonzero, let $\xi$ be a one-form, and
let
$$
\bear{rl}
X_{abcd}&=\xi_{(a}\xi_b\xi_c\xi_{d)_0} \\
&=\xi_a\xi_b\xi_c\xi_d \\
&\qquad-\frac18(\xi_a\xi_bg_{cd}
+\xi_a\xi_cg_{bd}+\xi_a\xi_dg_{bc}+\xi_b\xi_cg_{ad}+\xi_b\xi_dg_{ac}+
\xi_c\xi_dg_{ab})
|\xi|^2 \\
&\qquad+\frac1{48}(g_{ab}g_{cd}+g_{ac}g_{bd}+g_{ad}g_{bc})|\xi|^4,
\eear
$$
where $|\xi|^2=\xi^a\xi_a\,$. 
Direct calculation shows that
\beq\label{sixteenth}
(X\bullet Y)^{abcd}(X\bullet Y)_{abcd}=\tfrac1{16}|\xi|^8Y^{abcd}Y_{abcd}\,,
\eeq
where $|\xi|^2:=\xi^a\xi_a\,$.  This shows that $X\bullet Y$ is nonzero
if $\xi$ and $Y$ are.  
(This calculation is quite special to dimension 4; in higher dimensions,
the dependence on $\xi$ is not just through $|\xi|^2$.)

In fact, the computation of $X\bullet Y$ is exactly that of the
leading symbol of the operator $D$ of 
\nnn{block},
and \nnn{sixteenth}
shows that the leading symbol of $D^*D$ is $|\xi|^8/16$.
In other words,
$D^*D$ has principal part $\D^4/16$,
where $\D=-\nd^a\nd_a\,$.  

A more concrete workout of the duality issue can be obtained by writing
$X\bullet$ (for $X$ as just above)
in terms of the exterior and interior multiplication 
$\e(\xi)$ and $\i(\xi)$ of
differential forms by a one-form $\xi$.  
Here we use the fact that an algebraic
Weyl tensor density is (among other things) a 
($\L^2\otimes\L^2$)-density, and the fact that for $\xi$ is a one-form,
the Hodge $\star$ anticommutes with $\i(\xi)\e(\xi)-\e(\xi)\i(\xi)$. 

We now have a tensorial realization of the compression\newline
$\Proj_{\bun(\l\mp 2e_i)}\cdot|_{\bun(\l\pm 2e_i)}\,$, and may
conclude:

\begin{corollary}\label{weyltwo} The operator
\beq\label{weylalso}
\bear{rl}
Y^a{}_b{}^c{}_d\mapsto &
\{\N_{(a}\N_c\N_e\N_{f)_0}+10\V_{(ac}\N_e\N_{f)_0}
+10(\N_{(a}\V_{ce})\N_{f)_0} \\
&+3(\N_{(a}\N_c\V_{ef)_0})+9\V_{(ac}\V_{ef)_0}\}
Y^e{}_b{}^f{}_d \\
-&
\{\N_{(a}\N_d\N_e\N_{f)_0}+10\V_{(ad}\N_e\N_{f)_0}
+10(\N_{(a}\V_{de})\N_{f)_0} \\
&+3(\N_{(a}\N_d\V_{ef)_0})+9\V_{(ad}\V_{ef)_0}\}
Y^e{}_b{}^f{}_c \\
+&
\{\N_{(b}\N_d\N_e\N_{f)_0}+10\V_{(bd}\N_e\N_{f)_0}
+10(\N_{(b}\V_{de})\N_{f)_0} \\
&+3(\N_{(b}\N_d\V_{ef)_0})+9\V_{(bd}\V_{ef)_0}\}
Y^e{}_a{}^f{}_c \\
-&
\{\N_{(b}\N_c\N_e\N_{f)_0}+10\V_{(bc}\N_e\N_{f)_0}
+10(\N_{(b}\V_{ce})\N_{f)_0} \\
&+3(\N_{(b}\N_c\V_{ef)_0})+9\V_{(bc}\V_{ef)_0}\}
Y^e{}_a{}^f{}_d
\eear
\eeq
is conformally invariant 
$\cW^a{}_b{}^c{}_d$ to 
$\cW_{abcd}\,$, and carries the subbundle
$(\cW_\pm)^a{}_b{}^c{}_d$
to $(\cW_\mp)_{abcd}\,$.
\end{corollary}

There has also been some interest in tensorial realizations of
the first-order operators $D_i$ of \nnn{fourops}.  
Note that by the result of Fegan \cite{feg}, any SO$(n)$-invariant
first-order operator between irreducible SO$(n)$-bundles is a 
compression of the covariant derivative (i.e.\ has the form
${\rm Proj}_{\bun(\m)}\nd|_{\bun(\l)}$), and ``promotes'' to a
conformally covariant operator $\bun[w|\l]\to\bun[w-1|\m]$,
for a unique $w$ which is computable from $\l$ and $\m$.
With this in mind, our task in writing down the $D_i$ reduces
to writing nonzero 
SO$(n)$-invariant first-order operators that move between the
bundles advertised.

The first 
may be realized as a divergence:
$$
D_1:Y_{abcd}\mapsto\eta_{bcd}=\nd^aY_{abcd}\,.
$$
If we start in $\bun(2,2\e)$, where $\e=\pm 1$, this lands us in
the bundle $\bun(2,\e)$, which has a realization as the totally trace-free
tensors $\eta_{bcd}=\eta_{b[cd]}$ which have duality $\e$
in the $[cd]$
indices, and satisfy the Bianchi-like identity
$\eta_{bcd}+\eta_{cdb}+\eta_{dbc}=0$.  Let us denote 
this symmetry type (as a Riemannian bundle) by 
$\cA_{bcd}\cong_{{\rm SO}(n)}\bun(2,1)\oplus\bun(2,-1)$.  
We then switch to an alternative realization $\cA'_{abc}$ of 
$\bun(2,1)\oplus\bun(2,-1)$
as the totally trace-free 3-tensors $\eta'_{cab}=\eta'_{c(ab)}$ also 
satisfying
a Bianchi-like identity.  The SO$(n)$-equivariant map between the two
realizations is $\eta\mapsto\eta'$, where
$$
\eta'_{cab}:=\eta_{abc}+\eta_{bac}\,.
$$
(We have not bothered to normalize this map to an
isometry, as we are just working up to nonzero multiples.)
Applying a divergence in the first argument, we have $D_2$:
$$
\eta'_{cab}\mapsto\a_{ab}=\nd^c\eta'_{cab}\,;
$$
this lands us in $\cE_{(ab)_0}\,$.

To get (by $D_3$) to the bundle
$\bun(2,-\e)$, say $D_3\alpha=\beta$, we first take
\beq\label{comb1}
\bear{rl}
\beta''_{cab}:=&\frac23\nd_c\a_{ab}-\frac13\nd_a\a_{bc}
-\frac13\nd_b\a_{ca} \\
&\qquad-\frac19(g_{ca}\nd^e\a_{eb}+g_{cb}\nd^e\a_{ae})
+\frac29g_{ab}\nd^e\a_{ec}\,.
\eear
\eeq
This lands us in $\cA'_{cab}\,$;
To pick out the $\bun(2,-\e)$ summand, we go to the alternate realization
$\cA_{bcd}\,$:
\beq\label{comb2}
\b'_{cab}=\b''_{abc}-
\b''_{bac}\,.
\eeq
Let $\b_{cab}$ be the $(-\e)$-dual part of $\b'_{cab}$ in the $[ab]$ indices.
This process, $\a\mapsto\b''\mapsto
\b'\mapsto\b$, is the operator $D_3\,$.

Finally, we need $D_4$ to get us 
to the $(-\e)$-dual tensors with Weyl symmetry and trace type.
To accomplish this, we first take the map
\beq\label{comb3}
\b_{dab}\mapsto\bar{Z}_{cdab}:=\nd_{[c}\b_{d]ab}+\nd_{[a}\b_{b]cd}.
\eeq
The result of this process clearly satisfies the identities
$\bar{Z}_{cdab}=\bar{Z}_{cd[ab]}=\bar{Z}_{abcd}\,$, 
and a short computation shows
that in addition, $\bar{Z}_{cdab}+\bar{Z}_{cabd}+\bar{Z}_{cbda}=0$.  
Thus $\bar{Z}$ has 
curvature symmetries.
It is not, however, totally trace-free, though its double traces
$\bar{Z}^{ab}{}_{ab}$ do vanish (using the fact that $\b$ is totally 
trace-free).  The tensor
\beq\label{comb4}
Z_{cdab}:=\bar{Z}_{cdab}
-\frac12(\bar{Z}^e{}_{deb}g_{ca}+\bar{Z}_c{}^e{}_{ae}g_{db}
+\bar{Z}^e{}_{dae}g_{cb}+\bar{Z}_c{}^e{}_{eb}g_{da})
\eeq
is totally trace-free and enjoys curvature symmetries; i.e.\ it has
Weyl symmetry and trace type.
Since $\bun(1,0)\otimes\bun(2,-\e)$ has a $\bun(2,-2\e)$
summand but no $\bun(2,2\e)$ summand, $Z$ has duality $-\e$.

\subsection{Epilogue: BGG diagrams, other metric signatures, and
standard operators}\label{bggsection}\label{epilog}\label{standard}

BGG diagrams of tensors in 
4-dimensional Riemannian conformal
geometry are parameterized by similarity classes of integral
rho-shifted weights
$[[a|b,c]]$ which are strictly dominant after the bar.  (Recall \nnn{dom}
and the remarks immediately following.)  
Here we are speaking only of {\em tensorial} BGG diagrams; to
include those that depend on spin structure, we just need to 
admit properly half-integral $[[a|b,c]]$.  (See below for a discussion
of how the Dirac operator fits into this picture.)
{\em Regular} BGG diagrams correspond to similarity classes of
cardinality 6, and are in one-to-one correspondence with
triples $a,b,c$ of integers with $a>b>|c|$.  These appear as follows:
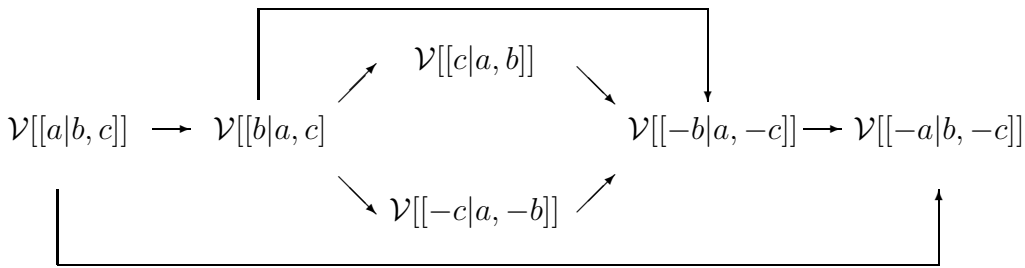
\setlength{\unitlength}{.95pt}
        \begin{center}\begin{picture}(436,100)(8,-55)

\put(28,-20){\line(0,-1){30}}     
\put(28,-50){\line(1,0){351}}
\put(379,-50){\vector(0,1){30}}

\put(8,0){$\bun[[a|b,c]]$}        

\put(90,0){$\bun[[b|a,c]$}      

\put(108,16){\line(0,1){36}}    
\put(108,52){\line(1,0){179}}
\put(287,52){\vector(0,-1){37}}

\put(170,28){$\bun[[c|a,b]]$}      
\put(160,-32){$\bun[[-c|a,-b]]$}   
\put(255,0){$\bun[[-b|a,-c]]$}     
\put(345,0){$\bun[[-a|b,-c]]$}     

\put(66,4){\vector(1,0){15}}       

\put(140,15){\vector(1,1){15}}     
\put(140,-15){\vector(1,-1){15}}   
\put(235,29){\vector(1,-1){15}}    
\put(235,-29){\vector(1,1){15}}    
\put(325,4){\vector(1,0){15}}      
\end{picture}\end{center}
\setlength{\unitlength}{1pt}

For example, the de Rham complex extends to a BGG diagram
with $a=2$, $b=1$, and $c=0$.
All compositions in this diagram vanish on $S^4$, except for one
linear combination of the two compositions around the diamond;
we represent this composition by the shorter rectangular arrow.
For the de Rham diagram, this surviving composition is the
Maxwell operator $d\star d$ on vector potentials, and the longest
arrow is the {\em Paneitz operator} mentioned above in Section
\ref{prelim}.

{\em Singular} BGG diagrams (for 4-dimensional conformal geometry)
correspond to similarity classes of cardinality 2.  Each gives rise to
a single (nonzero and non-identity) operator.  If $a>|c|$, we have an
operator $\bun[[a|a,c]]\to \bun[[-a|a,-c]]$, and if $a>c>0$, we have
the operators $\bun[[c|a,\pm c]]\to\bun[[-c|a,\mp c]]$.  (In the last
case the $\pm$ sign parameterizes two similarity classes.)  Our
operators on Weyl tensor densities, $\bun[[2|3,\pm
2]]\to\bun[[-2|3,\mp 2]]$, are of this final type.  In higher even
dimension $n$, the cardinality of a similarity class of bundles is
either $n+2$ (the regular case) or $2$ (the singular case).
A regular diagram just extends the 4-dimensional one above in the
obvious way, with conformal weights decreasing as one moves to the
right.

All operators have arbitrarily curved conformally invariant generalizations,
except for some of the longest arrows in regular diagrams.  For 
example, the Paneitz operator is conformally invariant in the arbitrarily
curved case, but the operator $\bun[[3|1,0]]\to\bun[[-3|1,0]]$
is known not to have an arbitrarily curved generalization \cite{crgnonex}.

All differential operators invariant under the conformal group of
round $S^n$ are captured in BGG diagrams (when one includes long arrows).
In particular, 
consider homogeneous combinations $D$ of
$\nd\cdots\nd$ terms whose index combinatorics are such that
$D:\bun(\l)\to\bun(\m)$ for some $\l,\m$ -- i.e.\ combinations that
pass between irreducible Riemannian bundles.
One might harbor the
naive hope that any such combination could be completed
to a conformally invariant differential operator by first assigning
appropriate conformal weights, and then adding lower-order terms. 
This must fail in general, since for a given $\nd\cdots\nd$ expression
to have any chance, it must be
(in the round $S^n$ case) the principal part of 
an operator in a BGG diagram.
If the expression passes this test, 
it may still fail in the conformally curved case, 
if its position in the round BGG was that of the longest arrow.

If we wish to speak of tensor-spinor bundles, we just need to add
bundles with proper half-integer entries to the above discussion.  For
example, the Dirac operator carries
$\bun[[\frac12|\frac32,\pm\frac12]]
\to\bun[[-\frac12|\frac32,\mp\frac12]]$, 
and so is much like
our Weyl tensor density operators.  The operator $\bun[[1|2,\pm
1]]\to\bun[[-1|2,\mp 1]]$ 
is the form-density operator of \cite{tbcpde} 
in the case of 2-forms in 4 dimensions; this interchanges the
two dualities:
$(\ce_\pm)_{[ab]}[1]\to (\ce_\mp)_{[ab]}[-1] $.

The operator of Theorem \ref{lambdathm} may occur in regular BGG
diagrams.  For example, one of the simplest operators we could construct
from the theorem carries scalar densities to trace-free symmetric 4-tensor
densities, $\bun[[5|1,0]]\to\bun[[1|5,0]]$; that is, $\cE[3]\to
\cE_{(abcd)_0}[3]$.  This is the first arrow in the BGG diagram above 
with $a=5$, $b=1$, $c=0$.
In tensor notation, the operator is
$$
\bear{rl}
f\mapsto & (\nd_{(a}\nd_b\nd_c\nd_{d)_0}+
10\V_{(ab}\nd_c\nd_{d)_0}+10(\nd_{(a}\V_{bc})\nd_{d)_0} \\
 &\qquad +3(\nd_{(a}\nd_b\V_{cd)_0})
+9\V_{(ab}\V_{cd)_0})f.
\eear
$$
We could also take trace-free symmetric 4-tensor densities to 
scalar densities: $\bun[[-1|5,0]]\to\bun[[-5|1,0]]$ or
$\cE_{(abcd)_0}[1]\to\cE[-7]$; this is in fact the formal adjoint
of the operator just above, and is also the final arrow in the same
BGG diagram.  A tensorial realization is
$$
\bear{rl}
\f_{abcd}\mapsto & (\nd^a\nd^b\nd^c\nd^d+
10\V^{ab}\nd^c\nd^d+10(\nd^a\V^{bc})\nd^d \\
&\qquad +3(\nd^a\nd^b\V^{cd})
+9\V^{ab}\V^{cd})\f_{abcd}\,.
\eear
$$  
In fact, this operator is contained in a class of 
conformally invariant operators, the
\begin{equation}\label{dnw}
\cE_{(a_1\ldots a_k)_0}[k-p-n+1]\to
\cE_{(b_1\ldots b_p)_0}[p-k-n+1]\ \ {\rm with}\ k>p,
\end{equation}
that plays a featured role in the recent work of Dolan, Nappi, and Witten
\cite{DNW}.

For any specified operator order $p$, \cite{goox} provides 
an analogue of Theorem
\ref{lambdathm} (where $p=4$), and an elementary proof along the
lines of that of Section \ref{stdfourth} above is possible.
Among other things, this allows one to write the lower-order terms of
the operators \nnn{dnw}.
The $p=1$ theorem is the result of
Fegan mentioned above.  The first of these involve
compressing the expressions 
$$
\bear{cc}
\nd  &  (p=1) \\
\nd\nd+\V &  (p=2) \\
\nd\nd\nd+4\V\nd+2(\nd\V) & (p=3) \\
\nd\nd\nd\nd+10\V\nd\nd+10(\nd\V)\nd+3(\nd\nd\V)+9\V\V\qquad & (p=4)
\eear
$$
Section 5 of \cite{goox} also gives the analogous expressions for $p=5,6,7$.
As in Theorem \ref{lambdathm}, these same expressions turn 
up in other dimensions \cite{Gothesis,CSSstandard}.
Things can be made to look more symmetric if we write expressions in
which terms act on everything to their right; for example
$\nd^4+4\nd\V\nd+3(\nd\nd\V+\V\nd\nd)+9\V\V$ for the fourth-order operator.

For example, back in dimension 4, we can get a conformally invariant operator
$\bun[[2|3,0]]\to\bun[[0|3,\pm 2]]$; that is, from trace-free symmetric
2-tensors with the index configuration $\a^a{}_b$ to 
$(\cW_\pm)^a{}_{bcd}\,$.
These operators appear in the {\em gravitational diagram}; that is,
the regular 4-dimensional BGG diagram with $a=3$, $b=2$, and $c=0$, and may
be interpreted as linearized Weyl curvature operators applicable to
a trace-free metric perturbation.  

In fact, the index combinatorics are given above in (\ref{comb1}-\ref{comb4}),
and the operators are the self-dual and anti-self-dual projections of
the expression 
(setting $\J:=\V^a{}_a$):
$$
\bear{rl}
\a_{ab}\mapsto &
\a_{ac|(bd)}-\a_{ad|(bc)}-\a_{bc|(ad)}+\a_{bd|(ac)} \\
+&\frac12g_{ac}(-\a_{bd|e}{}^e+\a_b{}^e{}_{|(de)}+\a_d{}^e{}_{|(be)}) 
-\frac12g_{ad}(-\a_{bc|e}{}^e+\a_b{}^e{}_{|(ce)}+\a_c{}^e{}_{|(be)}) \\
-&\frac12g_{bc}(-\a_{ad|e}{}^e+\a_a{}^e{}_{|(de)}+\a_d{}^e{}_{|(ae)})
+\frac12g_{bd}(-\a_{ac|e}{}^e+\a_a{}^e{}_{|(ce)}+\a_c{}^e{}_{|(ae)}) \\
+&\frac13(g_{ad}g_{bc}-g_{ac}g_{bd})\a^{ef}{}_{|ef} 
+\V_{bd}\a_{ac}-\V_{bc}\a_{ad}-\V_{ad}\a_{bc}+\V_{ac}\a_{bd} \\
+&\frac12g_{ac}(-\J\a_{bd}+\V_{de}\a_b{}^e+\V_{be}\a_d{}^e) 
-\frac12g_{ad}(-\J\a_{bc}+\V_{ce}\a_b{}^e+\V_{be}\a_c{}^e) \\
-&\frac12g_{bc}(-\J\a_{ad}+\V_{de}\a_a{}^e+\V_{ae}\a_d{}^e) 
+\frac12g_{bd}(-\J\a_{ac}+\V_{ce}\a_a{}^e+\V_{ae}\a_c{}^e) \\
+&\frac13(g_{ad}g_{bc}-g_{ac}g_{bd})\V_{ef}\a^{ef}.
\eear
$$
Going the other way, we can get operators $\bun[[0|3,\pm 2]]\to\bun[[-2|3,0]]$
by
$$
Y_{abcd}\mapsto\nd^b\nd^dY_{abcd}+\V^{bd}Y_{abcd}\,.
$$
Recalling the discussion of Section \ref{tens}, the self-dual and
anti-self-dual parts of
the Weyl tensor of the conformal structure live in 
$\bun[-2|2,\pm 2]=\bun[[0|3,\pm 2]]$, so these operators may be 
applied to $(C_\pm)^a{}_{bcd}\,$.  The result of applying to the
full Weyl tensor is called the {\em Bach tensor}:
\beq\label{bach}
\cB^a{}_c:=\nd^b\nd^dC^a{}_{bcd}+\V^{bd}C^a{}_{bcd}\,.
\eeq
In fact, by the uniqueness of the Bach tensor as a natural conformally
invariant section of $\cE_{(ab)_0}[-2]$ in dimension 4, together with
the universality of the calculation and the possibility of orientation
reversal, we must recover $\frac12\cB$ upon application of the above 
operator
to either of $C_\pm\,$.

Though we have proceeded throughout under the assumption of Riemannian
metric signature, the question of conformal invariance of abstract
index tensor expressions is signature
independent. 
Thus we also have a result, in dimension 4, for 
Lorentzian and signature $(2,2)$ conformal structures.
One only needs to note that the self-dual vs.\ anti-self-dual split
becomes, for Lorentz signature, an $\eye$-dual vs. $-\eye$-dual split.
(In general, on $p$-forms in dimension $n$ and a signature with
$q$ minus signs, $\star\star=(-1)^{p(n-p)+q}$.)

\newsavebox{\CROSS}
\savebox{\CROSS}[8pt]{\begin{picture}(8,8)(0,0)
                      \put(-5,-3){$\times$}
                      \end{picture}}

\newsavebox{\DISK}
\savebox{\DISK}[8pt]{\begin{picture}(8,8)(0,0) \put(-4.5,-3){$\bullet$}
                \end{picture}}

\newcommand{\ecross}{\begin{picture}(5,5)            
                     \put(1,0){\usebox{\CROSS}}
                     \end{picture}}
\newcommand{\cross}{\usebox{\CROSS}}                 
\newcommand{\disk}{\usebox{\DISK}}                   
\newcommand{\dynsup}[1]{\begin{picture}(8,8)(0,0)    
                       \put(-5,6){\makebox[8pt]{\scriptsize #1}}
                       \end{picture}}
\newcommand{\dynr}[1]{\begin{picture}(8,8)(0,0)    
                       \put(6,0){\makebox[8pt][l]{\scriptsize #1}}
                       \end{picture}}
\newcommand{\oxo}{\begin{picture}(40,10)(0,-4)
                  \put(4,0){\line(1,0){32}}
                  \put(4,0){\disk}
                  \put(20,0){\cross}
                  \put(36,0){\disk}
                  \end{picture}}

\newcommand{\fourdimn}[3]{\put(4,0){\dynsup{#1}}
                          \put(20,0){\dynsup{#2}}
                          \put(36,0){\dynsup{#3}}}

\newcommand{\oxon}[3]{\begin{picture}(45,13)(-2,0)  
                  \put(0,-4){\oxo}
                  \fourdimn{#1}{#2}{#3}
                  \end{picture}}

\newcommand{\Xoxo}{\begin{picture}(57,10)(0,-4)   
                 \put(4,0){\line(1,0){50}}
                 \put(4,0){\disk}
                 \put(28,0){\cross}
                 \put(54,0){\disk}
                 \end{picture}}

\newcommand{\Xoxon}[3]{\begin{picture}(66,10)(-5,0)
                 \put(0,-4){\Xoxo}
                 \put(4,0){\dynsup{#1}}
                 \put(28,0){\dynsup{#2}}
                 \put(54,0){\dynsup{#3}}
                 \end{picture}}

               In dimension 4 the operators in Theorem \ref{lambdathm}
               are essentially a subfamily of the so-called {\em standard
               operators} constructed in \cite{goox} (see also
               \cite{Eambi}).  There, on a complex holomophic
               conformal spin manifold $ \cal M$, conformally invariant
               operators are proliferated as direct images of a class
               of natural operators on the total space of the bundle
               of null directions of $ \cal M$. Once the operators are
               constructed in this way it is clear that the same
               formulae yield conformally invariant operators on a
               real conformal 4-manifold of any signature.  In
               \cite{goox}, irreducible holomorphic bundles are
               described in terms of weights on Dynkin diagrams as in
               $\co(\oxon{a}{b}{c}) $ and the order of a differential
               operator $\co(\oxon{a}{b}{c})\to \co(\Xoxon{d}{e}{f})$
               is the difference $ \frac{d+2e+f}{2}-\frac{a+2b+c}{2}$. 
               Note in particular that $ D_{abcd}$ from \cite{goox}
               yields the formula (\ref{projform}). From that source
               we see $ D_{abcd}$ will yield fourth order conformally
               invariant differential operators
               $\co(\oxon{a}{3}{c})\to \co(\Xoxon{a+4}{-5}{c+4})$,
               $\co(\oxon{a}{-a+2}{c})\to
               \co(\Xoxon{a-4}{-a-2}{c+4})$,   $\co(\oxon{a}{-c+2}{c})\to
               \co(\Xoxon{a+4}{-c-2}{c-4})$,
               $\co(\oxon{a}{1-a-c}{c})\to
               \co(\Xoxon{a-4}{1-a-c}{c-4})$ where the integers over
               the uncrossed nodes must be non-negative. For integers
               $ a,b,c$ with $ a,c$ non-negative, the representation $
               \oxon{a}{b}{c}$ corresponds to $[\frac{a+2b+c}{2}|
               \frac{a+c}{2},\frac{c-a}{2}]$ in our current notation.
               Thus these four classes of operator are respectively
               the operators $\bun[[\tl_1+2|\tl_1-2,\tl_2 ]]\to
               \bun[[\tl_1-2|\tl_1+2,\tl_2 ]]$,
               $\bun[[\tl_2+2|\tl_1,\tl_2-2 ]]\to
               \bun[[\tl_2-2|\tl_1,\tl_2+2 ]]$,
               $\bun[[-\tl_2+2|\tl_1,\tl_2+2 ]]\to
               \bun[[-\tl_2-2|\tl_1,\tl_2-2 ]]$, and,
               $\bun[[-\tl_1+2|\tl_1+2,\tl_2 ]]\to
               \bun[[-\tl_1-2|\tl_1-2,\tl_2 ]]$ of the theorem.

In other even dimensions the analogue of this construction
\cite{Gothesis} again yields all the operators of the theorem
including the formula \nn{projform}, but in odd dimensions the
operator is missed whenever it occurs as the middle operator in
the BGG pattern.

\end{document}